\documentclass[graybox]{svmult}

\usepackage{multirow}

\usepackage{mathptmx}       
\usepackage{helvet}         
\usepackage{courier}        
\usepackage{type1cm}        
\usepackage{makeidx}         
\usepackage{graphicx}        
\usepackage{multicol}        
\usepackage[bottom]{footmisc}

\makeindex             

\begin{document}

\title*{Measuring social spam and the effect of bots on information diffusion in social media}
\author{Emilio Ferrara}
\institute{Emilio Ferrara \at University of Southern California, Information Sciences Institute; 4676 Admiralty way \#1001, Marina del Rey, CA (USA); \email{emiliofe@usc.edu}}
%
%
\maketitle

\abstract*{Bots have been playing a crucial role in online platform ecosystems, as efficient and automatic tools to generate content and diffuse information to the social media human population. In this chapter, we will discuss the role of social bots in content spreading dynamics in social media. In particular, we will first investigate some differences between diffusion dynamics of content generated by bots, as opposed to humans, in the context of political communication, then study the characteristics of bots behind the diffusion dynamics of social media spam campaigns.}

\abstract{Bots have been playing a crucial role in online platform ecosystems, as efficient and automatic tools to generate content and diffuse information to the social media human population. In this chapter, we will discuss the role of social bots in content spreading dynamics in social media. In particular, we will first investigate some differences between diffusion dynamics of content generated by bots, as opposed to humans, in the context of political communication, then study the characteristics of bots behind the diffusion dynamics of social media spam campaigns.}

\section{Introduction} \label{sec:introduction}
Social media have received widespread recognition as enablers of modern society communication~\cite{lazer2009life, kwak2010twitter, cha2010measuring, boyd2012critical, kumpel2015news}, as a tool to democratize discussion about politics~\cite{adamic2005political, effing2011social, bekafigo2013tweets, carlisle2013social, digrazia2013more, lutz2014beyond, yang2016social} and social issues~\cite{gonzalez2011dynamics, gonzalez2013broadcasters, conover2013geospatial, conover2013digital, varol2014evolution, barbera2015critical, theocharis2015using}, and even as an effective system to respond to crises and emergencies~\cite{sutton2008backchannels, yates2011emergency, gao2011harnessing, yin2012using, latonero2013emergency}. 

The benefits of the rise to popularity of social media are hard to quantify, as they touch billions of people every day, all over the world. However, as early as 2006, concerns have been raised regarding the possibility of manipulating public opinion through social media~\cite{howard2006new}. Particularly problematic can be the fact that social media have proved effective in influencing individuals, their believes and behaviors~\cite{aral2011creating, centola2011experimental, kramer2014experimental, ferrara2015measuring, monsted2017evidence}.
These concerns have been later proved well grounded by several scientific studies, which highlighted a variety of manipulation strategies and related contexts where such forms of abuse can take place~\cite{ratkiewicz2011truthy, metaxas2012social, el2013social, ferrara2015manipulation, howard2016bots, shorey2016automation, Varol2017, FM8005}. 

One way to manipulate social media is by using social bots, algorithmically-controlled accounts that emulate the activity of human users but operate at much higher pace (e.g., automatically producing content or engaging in social interactions), while successfully keeping their robotic identity undisclosed~\cite{hwang2012socialbots, messias2013you, ferrara2016rise, ICWSM1715587}.

Evidence of the adoption of social media bots to attempt manipulating political communication dates back nearly a decade: during the 2010 U.S. midterm elections, social bots were employed to support some candidates and smear others, by injecting thousands of tweets pointing to websites with fake news~\cite{ratkiewicz2011detecting}. The research community reported another similar case around the time of the 2010 Massachusetts special election~\cite{metaxas2012social}. Campaigns of this type are sometimes referred to as astroturf or Twitter bombs. Unfortunately, most of the times, it has proven impossible to determine who's behind these types of operations~\cite{ferrara2016rise, kollanyi2016bots, bessi2016social}.
Governments, organizations, and other entities with sufficient resources, can obtain the technological capabilities to deploy thousands of social bots and use them to their advantage, either to support or to attack particular political figures or candidates. 

Bots have been used in other contexts too, most prominently for social spamming and social phishing purposes~\cite{jagatic2007social, thomas2011suspended, song2011spam, jin2011data, yang2012analyzing,  mukherjee2012spotting, thomas2013trafficking}. A large body of scientific literature covers the challenges related to detecting social spam~\cite{markines2009social, gao2010detecting, zhang2012detecting}, spam bots~\cite{lee2010social, lee2010uncovering, stringhini2010detecting, boshmaf2011socialbot}, fake reviews~\cite{mukherjee2012spotting}, etc. Differently from traditional Internet spam, distributed via email or mailing lists, social spam proliferates in online platforms, and bots have been extensively used to make its diffusion more effective. Although much work has been devoted to characterize and detect social spam campaigns or spam bots, the interplay between these two, and in particular the effect of spam bots on the diffusion of spam in social media, has not received much attention.

\subsection*{Contributions of this chapter}
This chapter aims at investigating both the directions of social bots influence on political discussion, and spam bots influence in social spam campaigns. In particular, we will be concerned with measuring the role and effects of bots in social media information spreading dynamics. The scope and contributions of this chapter are therefore threefold: 

\begin{itemize}
\item We will first review how social bots, and in particular Twitter bots, are created, how they operate, and what are the challenges in detecting them (see Section~\S\ref{sec:bots}). The literature discussed here will be mostly aligned with a recent review paper we published on \textit{Communications of the ACM}~\cite{ferrara2016rise}.

\item We will then discuss how social bots have been used during the 2016 US Presidential Election to sway the discussion around the presidential candidates, and to frame agendas and messages attaching particular sentiments. This review (see Section~\S\ref{sec:election}) will be based on results we recently published~\cite{bessi2016social}.

\item Then, we will propose novel analysis of the effects of social spam bots on the diffusion of social spam campaigns and promotional content on Twitter (see Section~\S\ref{sec:spam}). We will investigate the differences between traditional spammers and social spam bots, provide a characterization of their most typical features, and describe their effect of the diffusion of social spam on Twitter.
\end{itemize}

\section{What Social Bots are and How They Operate}\label{sec:bots}

\subsection*{How to create a social spam bot}
In the early days of online social media, over one decade ago, creating a bot was not a simple task: a skilled programmer would need to sift through various platforms' documentation to create a software capable of automatically interfacing with the platform and operate functions in a human-like manner. For example, in 2009, we spent significant amounts of efforts to create a simple bot that would navigate Facebook pages and extract basic publicly-available social network information~\cite{catanese2011crawling}: that required the application of sophisticated Web scripting techniques~\cite{ferrara2014web} in conjunction with a trial-and-error approach to deal with the Web platform infrastructure. Similar efforts have been reported for other such type of early endeavors~\cite{coburn2011realboy,aiello2013people}

These days, the landscape has completely changed: indeed, it has become increasingly simpler to deploy social bots, so that, in some cases, no coding skills are required to setup accounts that perform simple automated activities: tech blogs often post tutorials and ready-to-go tools for this purposes. Various source codes for sophisticated social media bots can be found online as well, ready to be customized and optimized by the more technically-savvy users~\cite{kollanyi2016bots}. 

We inspected same of the readily-available Twitter bot-making tools and this is a (non-comprehensive) list of capabilities they provide: 

\begin{itemize}
\item Search Twitter for phrases/hashtags/keywords and automatically retweet them; 
\item Automatically reply to tweets that meet a certain criteria; 
\item Automatically follow any users that tweet something with a specific hashtag, keyword, or phrase; 
\item Automatically follow back any users that have followed the bot; 
\item Automatically follow any users that follow a specified user; 
\item Automatically add users tweeting about something to public lists; 
\item Search Google (and other engines) for articles/news according to specific criteria and post them, or link them in automatic replies to other users; 
\item Automatically aggregating public sentiment on certain topics of discussion; 
\item Buffer and post tweets automatically. 
\end{itemize}

Most of these bots can run within cloud services or infrastructures like Amazon Web Services (AWS) or Heroku, making it more difficult to block them when they violate the Terms of Service of the platform where they are deployed. 

Finally, a very recent trend is that of providing Bot-As-A-Service (BaaS): companies like RoboLike\footnote{RoboLike: \url{https://robolike.com/}} provide ``Easy-to-use Instagram/Twitter auto bots'' performing certain automatic activities for a monthly price. Advanced conversational bots powered by sophisticated Artificial Intelligence are provided by companies like ChatBots.io that allow anyone to ``Add a bot to services like Twitter, Hubot, Facebook, Skype, Twilio, and more''.\footnote{Pandora bot: \url{https://developer.pandorabots.com/}}

\subsection*{How to detect social bots}
The detection of social bots in online social media platform has proven a challenging task. For this reason, it has attracted a lot of attention from the computing research community. Even DARPA became interested to the point that a DARPA Challenge was organized, namely the 2016 DARPA Twitter Bot Detection~\cite{subrahmanian2016darpa}: over one dozen academic and industry teams participated, with University of Maryland, University of Southern California, and Indiana University topping the challenge.

For these reasons, the literature on social bot detection has become very extensive. We tried to summarize the most relevant approaches in a survey paper recently appeared on \textit{Communications of the ACM}~\cite{ferrara2016rise}: we refer the interested reader to that review for a deeper analysis of this problem.

In our review, we proposed a simple taxonomy to divide the social bot detection approaches proposed in literature into three classes: \emph{(i)} bot detection systems based on social network information; \emph{(ii)} system based on crowd-sourcing and leveraging human intelligence; \emph{(iii)} machine learning methods based on the identification of highly-revealing features that discriminate between bots and humans. In the following, we report some examples of these three classes.

\subsubsection*{Graph-based social bot detection.} 
Social bot detection has been framed as an adversarial setting~\cite{alvisi2013sok}: an adversary may control multiple social bots to impersonate different identities and infiltrate a system. Proposed detection strategies often rely on examining the structure of a social graph, and assume that bot accounts exhibit a small number of links to legitimate users, connecting mostly to other bots. This feature is exploited to identify densely interconnected groups of bots. Yet, a wise attacker may counterfeit the connectivity of the controlled bot accounts; this strategy would make the attack invisible to thse detection methods. 
To address this shortcoming, some systems also employ the paradigm of \emph{innocent by association}: an account interacting with a legitimate user is considered itself legitimate. Unfortunately, the effectiveness of such detection strategies is bound by the behavioral assumption that legitimate users refuse to interact with unknown accounts. This was proven unrealistic by various experiments~\cite{stringhini2010detecting,boshmaf2013design,elyashar2013homing}. On other platforms like Twitter and Tumblr, connecting and interacting with strangers is one of the main features. In these circumstances, the innocent-by-association paradigm yields high false positive rates. Moreover, real-world platforms may contain many mixed groups of legitimate users who fell prey of some bots~\cite{alvisi2013sok}, and sophisticated bots may succeed in large-scale infiltration making it impossible to detect them solely from network structure information. 
Despite its high false-positive rate, social network information can complement other sources of information to improve prediction accuracy, as demonstrated by prior work~\cite{ferrara2016rise}.

\subsubsection*{Crowd-sourcing social bot detection}
Some authors suggested  crowd-sourcing social bot detection, assuming that it would be a simple task for humans to evaluate an account's behavior and to observe emerging patterns and anomalies associated with bots~\cite{wang2012social}. Using data from Facebook and Renren (a popular Chinese online social network), the authors tested the efficacy of human detectors, using both expert annotators and workers hired online.
Although this strategy exhibited a near-zero false positive rate, it has proven unfeasible for several reasons: for existing  platform with large user bases, like Facebook and Twitter, manually verify millions of suspicious accounts has a prohibitive cost; even if large social network companies could afford to hire teams of analysts for this purpose~\cite{stein2011facebook}, such cost might not be sustainable for small social networks in their early stages; finally, exposing personal information to online workers for annotation would raise privacy issue~\cite{elovici2013ethical}. 

\subsubsection*{Feature-based social bot detection}
Encoding behavioral patterns into features, in conjunction with machine learning techniques to learn the signature of human and bot behavior, may be the most popular bot detection strategy. One example of feature-based system is represented by \emph{Bot or Not}: released in 2014, and constantly updated, this was the  first Twitter bot detection tool to be made publicly available~\cite{davis2016botornot}.\footnote{\url{http://truthy.indiana.edu/botornot}}
\emph{Bot or Not} implements a detection algorithm relying upon highly-predictive features capturing a variety of suspicious behaviors to separate social bots from humans. 
The system employs off-the-shelf supervised learning algorithms trained with examples of both humans and bots behaviors.
In addition to the classification results, \emph{Bot or Not} provides a variety of interactive visualizations that yield insights on the features exploited by the system. We will later describe how we used \textit{Bot or Not} for our studies.

Bots are continuously changing and evolving: 
the analysis of the highly-predictive behaviors that feature-based detection systems can detect may reveal interesting patterns and provide unique opportunities to understand how to discriminate between bots and humans.
User meta-data are considered among the most predictive features and the most interpretable ones~\cite{hwang2012socialbots,wang2012social}: we can suggest few rules of thumb to infer whether an account is likely a bot, by comparing its meta-data with that of legitimate users.
Further work, however, will be needed to detect sophisticated strategies exhibiting a mixture of humans and social bots features (sometimes referred to as \emph{cyborgs}). Detecting these  bots, or hacked accounts~\cite{zangerle2014hacked}, is currently impossible for feature-based systems. Recent studies suggested that some advanced social bots may no longer aim at mimicking human behavior, but rather at misdirecting attention to irrelevant information~\cite{abokhodair2014dissecting}: such \emph{smoke screening} strategies, requiring high degree of coordination among bots, can also escape feature-based detection systems.

\section{Applications and Case Studies}
In the following, we present two case studies. We first study the use of social bots in the context of the 2016 US Presidential Election (\textit{cf.} Section~\S\ref{sec:election}). The results we present are based on recently published work~\cite{bessi2016social}. Then, we discuss new results on the effect of bots on the diffusion of social media spam (\textit{cf.} Section~\S\ref{sec:spam}).

\subsection{Case study 1: Political campaigns}\label{sec:election}
In the introduction of this chapter, we discussed at length the widespread abuse of social media platforms. In the context of political campaigns, one could try to boost the popularity of a candidate, for example by creating the impression that there is an organic support behind that candidate; however, the apparent support can be artificially generated by means of orchestrated campaigns.
This phenomonon is commonly referred to as \textit{astroturf}, and it has long-lasting roots, starting from offline campaigns~\cite{lyon2004astroturf}, and evolving, during more recent times, into various forms of Internet~\cite{klotz2007internet} and social media~\cite{ratkiewicz2011truthy} campaigns. 
We report our study of social media astroturf in the context of the 2016 US Presidential Election next, with a special focus on the role of social bots. We discuss data collection first, then we go over the employed bot detection and sentiment analysis approaches. The case study concludes with some discussion of the insights our analysis yielded.

\subsubsection*{Data Collection} 
We manually crafted a list of hashtags and keywords related to the 2016 US Presidential Election. The list was compiled so that to contain a roughly equal number of hashtags/keywords associated with each major presidential candidate: we selected 23 terms in total, including 5 terms specifically for the Republican Party nominee Donald Trump (\#donaldtrump, \#trump2016, \#neverhillary, \#trumppence16, \#trump), 4 terms for the Democratic Party nominee Hillary Clinton (\#hillaryclinton, \#imwithher, \#nevertrump, \#hillary), and several terms relative to the four presidential debates. The full list of search terms is reported in our paper~\cite{bessi2016social}. 
By querying the Twitter Search API at regular intervals of 10 seconds, continuously and without interruptions in three periods between September 16 and October 21, 2016, we collected a large dataset constituted by 20.7 million tweets posted by nearly 2.8 million distinct users. We used the Twitter Search API\footnote{Twitter Search API: \url{https://dev.twitter.com/rest/public/search}} to obtain all tweets that contain the search terms, posted during the data collection period, rather than a sample of unfiltered tweets: this avoids incurring in the issues reported in the literature related to collecting sample data from the Twitter Stream API\footnote{Twitter Stream API: \url{https://dev.twitter.com/streaming/overview}} instead~\cite{morstatter2013sample}.

\subsubsection*{Bot detection}
Determining whether either human or a bot controls a social media account has proven a very challenging task~\cite{ferrara2016rise, subrahmanian2016darpa}. Our prior efforts produced an openly accessible solution called Bot Or Not~\cite{davis2016botornot}, consisting of a Python API\footnote{Bot or Not Python API: \url{https://github.com/truthy/botornot-python}} and a Website.\footnote{Bot or Not Website: \url{https://truthy.indiana.edu/botornot/}} 
As we briefly discussed earlier, Bot Or Not is a machine-learning framework that extracts and analyses a set of over one thousand features, spanning content and network structure, temporal activity, user profile data, and sentiment analysis to produce a score that suggests the likelihood that the inspected account is indeed a social bot. Extensive analysis revealed that the two most important classes of feature to detect bots are, maybe unsurprisingly, the metadata and usage statistics associated with the user accounts. 

The following indicators provide the strongest signals to separate bots from humans: (i) whether the public Twitter profile looks like the default one or it is customized (it requires some human efforts to customize the profile, therefore bots are more likely to exhibit the default profile setting); (ii) absence of geographical metadata (humans often use smartphones and the Twitter iPhone/Android App, which records as digital footprint the physical location of the mobile device); (iii) and activity statistics such as total number of tweets and frequency of posting (bots exhibit incessant activity and excessive amounts of tweets), proportion of retweets over original tweets (bots retweet contents much more frequently than generating new tweets), proportion of followers over followees (bots usually have less followers and more followees), account creation date (bots are more likely to have recently-created accounts), randomness of the username (bots are likely to have randomly-generated usernames).  We point the reader interested in further technical details to our prior work~\cite{ferrara2016rise, davis2016botornot}.

Bot Or Not has been trained with thousands of instances of social bots, from simple to sophisticated, and an accuracy of above 95\%~\cite{davis2016botornot}. Typically, Bot Or Not yields likelihood scores above fifty percent only for accounts that look suspicious to a scrupulous analysis. We adopted the Python Bot Or Not API to systematically inspect the most active users in our dataset. The Python Bot Or Not API queries the Twitter API to extract the most 300 tweets and all the publicly available account metadata, and feed this features to an ensemble of machine learning classifiers, which produce a bot score. To label accounts as bots, we use the fifty-percent threshold---which has proven effective in prior studies~\cite{ferrara2016rise, davis2016botornot}---an account is considered to be a bot if the bot score is above 0.5. 

Since the Python Bot Or Not API incurs in the query limitations imposed by the Twitter API, it would have been impossible to test all the 2.78 million accounts. Therefore, we tested the top 50 thousand accounts ranked by activity volume. Although these top 50 thousand users account for roughly only 2\% of the entire population, it is worth noting that they are responsible for producing over 12.6 million tweets, which is about 60\% of the total conversation. This choice gives us sufficient statistical power to extrapolate the distribution of bots and humans for the entire population without the need to test accounts that are only marginally involved in the conversation. 
Out of the top 50 thousand accounts, Bot Or Not assigned a bot score greater than the established 0.5 threshold, and therefore classified as likely bots, to a total of 7,183 users, responsible for 2,330,252 tweets. A total of 40,163 users (responsible for 10.3 million tweets) were labeled as humans. Bot Or Not labeled the remainder 2,654 users as unknown/undecided, either because their scores does not significantly diverge from the classification threshold of 0.5, or because the accounts have been suspended/deleted. Even if all the 2,654 users were bots, and Twitter suspended their accounts for violating the terms of service, this would suggest that roughly 70\% of the total bot population (the remainder 7,183 accounts) was still active on the platform at the time of our verification.
By extrapolating for the entire population, we estimate the presence of at least 400 thousand bots, accounting for roughly 15\% of the total Twitter population active in the U.S. presidential election discussion, and responsible for about 3.8 million tweets, roughly 19\% of the total volume. Additional statistics are summarized in our paper~\cite{bessi2016social}.

\subsubsection*{Sentiment analysis}
To understand how bots and humans discuss about the presidential candidates we will rely upon sentiment analysis. To attach a sentiment score to the tweets in our dataset, we used SentiStrength~\cite{thelwall2010sentiment}. SentiStrength is a sentiment analysis algorithm which has been specifically designed to annotate social media data. This design choice provides some desirable advantages: first, it is optimized to annotate short, informal texts, like tweets, that contain abbreviations, slang, and other non-orthodox language features; second, SentiStrength employs additional linguistic rules for negations, amplifications, booster words, emoticons, spelling corrections, etc. Applications of SentiStrength to social media data found it particularly effective at capturing positive and negative emotions with, respectively, 60.6\% and 72.8\% accuracy~\cite{thelwall2013heart}. We tested it extensively and also used it in prior studies to validate the effect of sentiment on the diffusion of information in social media~\cite{ferrara2015measuring}.
The algorithm assigns to each tweet $t$ a positive $P^+(t)$ and negative $P^-(t)$ polarity score, both ranging between 1 (neutral) and 5 (strongly positive/negative). Starting from the polarity scores, we capture the emotional dimension of each tweet $t$ with one single measure, the sentiment score $S(t)$, defined as the difference between positive and negative polarity scores: $S(t) = P^+(t) - P^-(t)$.
The above-defined score ranges between -4 and +4. The negative extreme indicates an strongly negative tweet, and occurs when $P^+(t)=1$ and $P^-(t)=5$. Vice-versa, the positive extreme identifies a strongly positive tweet labeled with $P^+(t)=5$ and $P^-(t)=1$. In the case $P^+(t)=P^-(t)$---positive and negative sentiment scores for a tweet $t$ are the same---the sentiment $S(t)=0$ of tweet $t$ is considered as neutral (note that the neutral class represent the majority, by construction, since it contains all tweets that have equal number of positive and negative words, as well as all tweets with no sentiment-labeled terms).

\begin{figure}[t]
\sidecaption[t]
\includegraphics[width=\columnwidth]{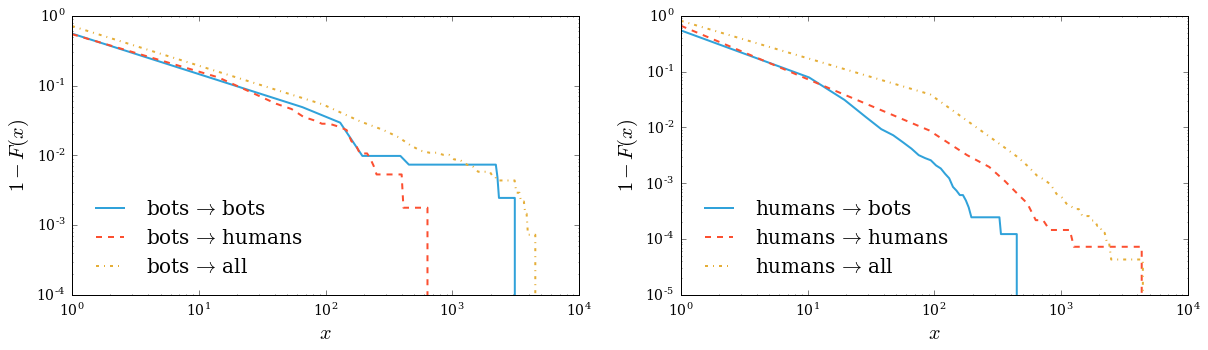}
\caption{Complementary cumulative distribution function (CCDF) of replies interactions generated by bots (left) and humans (right) (published in Bessi \& Ferrara, 2016~\cite{bessi2016social}).}
\label{fig:321}       
\end{figure}

\begin{figure}[t]
\sidecaption[t]
\includegraphics[width=\columnwidth]{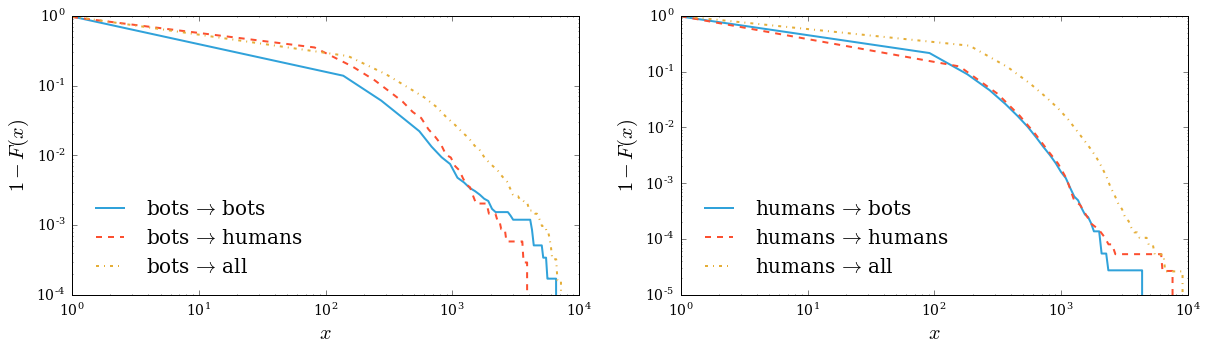}
\caption{Complementary cumulative distribution function (CCDF) of retweets interactions generated by bots (left) and humans (right) (published in Bessi \& Ferrara, 2016~\cite{bessi2016social}).}
\label{fig:322}       
\end{figure}

\subsubsection*{Partisanship and Supporting Activity}
We next inferred the partisanship of the users in our dataset. We used the 5 Trump-supporting hashtags (\#donaldtrump, \#trump2016, \#neverhillary, \#trumppence16, \#trump) and the 4 Clinton-supporting (\#hillaryclinton, \#imwithher, \#nevertrump, \#hillary) to attribute partisanships. In detail, we employed a simple heuristics based on hashtag adoption: for each user, we calculated the top 10 hashtags that appear in the tweets posted by that user. If the majority of hashtags support one particular candidate, we assigned the given user to that political faction (Clinton- or Trump-supporter). This is a very strict and conservative partisanship assignment, likely less prone to misclassification that may be yield by automatic machine-learning techniques not based on manual validation, e.g.,~\cite{conover2011political}. Our procedure yielded a small, high-confidence, annotated dataset constituted by 7,112 Clinton supporters (590 bots and 6,522 humans) and 17,202 Trump supporters (1,867 bots and 15,335 humans).

\subsubsection*{Analytic insight 1: Human vs. Bot engagement}

Figure~\ref{fig:321} and Figure~\ref{fig:322} show the Complementary Cumulative Distribution Functions (CCDFs) of the interactions respectively replies and retweets, initiated by bot and human users. Each plot disaggregates the interactions in three categories: (i) within group (for example bot-bot, or human-human); (ii) across groups (e.g., bot-human, or human-bot); and, (iii) total (i.e., bot-all and human-all). Both figures exhibit broad distributions typical of social media activity. What interestingly emerges from contrasting the two figures, is that humans are engaging in replies interactions significantly more (one order of magnitude difference) with other humans than with bots (see right panel of Figure~\ref{fig:321}). Conversely, bots fail to substantially engage humans and end up interacting via replies with other bots significantly more than with humans. Given that bots by design are intended to engage in interactions with humans, our observation goes against what we would have intuitively expected---similar paradoxes have been highlighted in our prior work~\cite{ferrara2016rise}.  One intuitive explanation to this phenomenon is that bots that are not sophisticated enough, cannot produce engaging-enough questions to foster meaningful discussions with humans. Figure~\ref{fig:322}, however, demonstrates that rebroadcasting is a much more effective channel of information spreading: there is no significant difference in the amounts of retweets that humans generate by rebroadcasting content produced by other humans or by bots. In fact, humans and bots retweet each other substantially at the same rate. This suggests that bots are being very effective at spreading information in the human population, which could have some nefarious consequences in the cases when humans fail at verifying the correctness and accuracy of such information and information sources.

\subsubsection*{Analytic insight 2: Human vs. Bot sentiment}

\begin{figure}[t]
\sidecaption[t]
\includegraphics[width=\columnwidth]{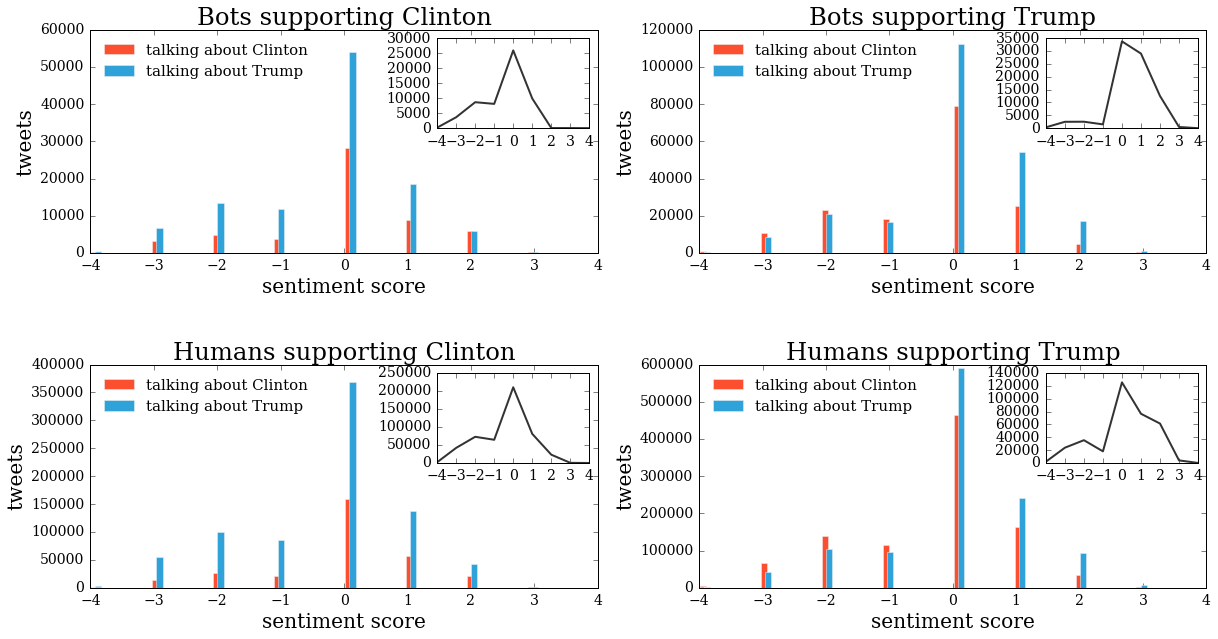}
\caption{Distributions of the sentiment of bots (top) and humans (bottom) supporting the two presidential candidates. The main histograms show the disaggregated volumes of tweets talking about the two candidates separately, while the insets show the absolute value of the difference between them (published in Bessi \& Ferrara, 2016~\cite{bessi2016social}).}
\label{fig:323}       
\end{figure}

To further understand how social media users (both bots and humans) are talking about the two presidential candidates, we explore the sentiment that the tweets convey. To this purpose, we rely upon sentiment analysis and in particular on \textit{SentiStrength}. Figure~\ref{fig:323} shows four panels: the top two panels illustrate the sentiment of the tweets produced by the bots, while the bottom two panels show the same information for tweets generated by humans. Furthermore, the two left panels show the support to Hillary Clinton (respectively by bots and humans), whereas the two right panel show the support to Donald Trump (respectively by bots and humans). The main histograms in each panel show the volume of tweets about Clinton or Trump, separately, whereas the insets show the difference between the two (this to illustrate the disproportion in support of the candidate of one's factions, as opposed to the other candidate).
What appears evident from contrasting the left and right panels is that, on average, the tweets produced by Trump's supporters are significantly more positive than that of Clinton's supporters, regardless of whether the source is human or bot. If we focus on Trump's bot supporters, we note that they generate almost no negative tweets; they indeed produce the most positive set of tweets in the entire dataset---a very significant fraction of these non-negative bot-generated tweets (about 200,000 or nearly two-third of the total) are in support of Donald Trump. This generates a stream of support that is at staggering odds with respect to the overall negative tone that characterizes the 2016 presidential election campaigns. The fact that bots produce systematically more positive content in support of a candidate can bias the perception of the individuals exposed to it, suggesting that there exists an organic, grassroots support for a given candidate, while in reality it is all artificially generated.
Some interesting insights emerge also from the analysis of Clinton's supporters: on average, human-generated tweets show slightly more positive sentiment toward the candidate than the bot-generated ones. Overall, a more natural distribution of tweets' sentiment emerges from the two groups of bots and human supporters, with a roughly equal number of positive and negative tweets being present in the pro-Clinton discussion. 
To further understand these dynamics, we manually analyzed two hashtags, namely \#NeverTrump and \#NeverHillary, as emblematic examples of campaigns explicitly devoted to target the candidate of one's opposing political leaning. The hashtag \#NeverTrump, used by supporters of the Democratic Candidate Hillary Clinton, accrued 105,906 positive tweets, and 118,661 negative ones, roughly an equal split; on the other hand, the hashtag \#NeverHillary pushed by Trump's supporters generated significantly more negative tweets (204,418) than positive ones (171,877). The paper~\cite{bessi2016social} reports various examples of tweets generated by bots, and the candidate they support.
A final consideration emerges when contrasting the pro-Clinton and pro-Trump factions: the former focuses much more on their candidate, with a significant number of tweets referring to Clinton. Conversely, pro-Trump supporters (humans and bots) devote a significant number of tweets to their opponent: in fact, the majority of negative tweets generated by both humans and bots are addressing Hillary Clinton. 

\subsection{Case study 2: Social Spam Campaigns}\label{sec:spam}
In the second part of this chapter, we study social spam campaigns. The widespread use of social media makes them an ideal target as a vector to diffuse spam campaigns. Indeed, spam has evolved, moving away from traditional vectors like emails and mailinglists~\cite{heymann2007fighting}, due to the increasing effectiveness of email spam filters, and migrating to social platforms like social media~\cite{gao2010detecting, chu2012detecting, zhang2012detecting} and  digital marketplaces~\cite{jindal2007review, pang2008opinion, mayzlin2014promotional}, etc. In the former scenario, the use of bots has been documented to generate artificial promotional campaigns, to advertise dubious products (whose sale is sometimes illicit), etc. In the latter, bots are exploited to generate and diffuse fake product reviews.
Next, we study social media spam, focusing on the effects of social bots in the diffusion of spam campaigns on Twitter. We first discuss social spam data collection, then introduce a tool named \textit{dynamical activity-connectivity map} we recently proposed to study the mechanisms of influence in social media. We conclude studying spam campaigns' sentiment and its interplay with bots' efficacy.

\subsubsection*{Data Collection}
Similarly to the political discussion scenario, we manually crafted a list of hashtags and keywords to collect our data. We focused on the tobacco-related discussion, and in particular electronic cigarettes. We identified this case study by noticing how spam seems to be a pervasive presence in this topic of discussion on Twitter~\cite{allem2016importance}. The list included over one hundred terms covering nicotine-related products (e.g., \textit{tobacco}, \textit{cigar}, \textit{cigarettes}, etc.),  electronic cigarettes (multiple variants like \textit{ecig}, \textit{e-cig}, \textit{ecigs}, \textit{e-cigs}, \textit{e-cigarette}, \textit{ecigarette}, etc.), vaping products (e.g., \textit{vape}, \textit{ehookah}, \textit{ejuices}, \textit{eliquids}, etc.), popular vaping brands (e.g., \textit{green smoke}, \textit{eversmoke}, etc.), health-related terms (e.g., \textit{second-hand smoke}, \textit{second-hand vape}), health campaigns terms (e.g., \textit{still blowing smoke}, \textit{not blowing smoke}, \textit{tobacco free kids}, etc.), and more.
We queried the Search API at regular intervals  from January 1 to September 30, 2015 and collected a large dataset constituted by over 9 million unique tweets.

\subsubsection*{Spam Detection}
Detecting social spam has proven a challenging and tedious task. 
The lack of a rigorous definition of what spam is makes  detection a complex problem. Although various detection techniques have been proposed in the machine learning literature, they carry some limitations: they are either outdated, being trained and tested on early (2008-2010) Twitter spam data~\cite{lee2010social, lee2010uncovering, stringhini2010detecting, boshmaf2011socialbot}, or overly-specific to detect certain types of  campaigns~\cite{markines2009social, gao2010detecting, zhang2012detecting, ferrara2016detection}. The first limitation becomes a problem due to the fact that bots evolve, becoming increasingly sophisticated thus rendering detection  less effective if training data is not current; the latter issue hinders the applicability of detection systems to a broader range of problem domains.

For the reasons above, to detect spam campaigns in our data and separate legitimate tobacco-related discussion from social spam, we implemented a novel strategy. We first performed traditional data cleaning operations on the texts of the tweets in our dataset, namely removing stop-words and punctuation, then tokenizing and stemming the terms. Afterwards, we elaborated the following iterative 3-stages detection procedure: 

\begin{enumerate}
\item We generated a list of keywords appearing in the tweets, ranked by frequency. 
\item Then, two independent human annotators manually identified and labeled keywords associated to spam campaigns appearing in the list of the top 250 most common keywords (to provide contextual information, the annotators had access to the full text of some example tweets where such keywords occur). 
\item Finally, all tweets containing spam-associated keywords are moved into a separate repository that we will call \textit{spam dataset}; the iterative process then restarts. It is worth noting that, at each next iteration of the algorithm, the ranked list of keywords changes because the spam keywords identified at stage 2 are removed. 
\end{enumerate}

The process ended when the list of top 250 most common keywords did not contain any spam-associated term.  This yielded a manually-curated list of 87 spam keywords,\footnote{The combination of the top 250 non-spam keywords, plus the 87 spam keywords, accounts for over 90\% of all tweets in the original dataset.} that appear in the \textit{spam dataset} accounting for 3.06M unique tweets posted by over 850 thousand distinct users. Of these users, about 74K posted more than one tweet. We will focus our attention, for the rest of our analysis, on these 74K active spammers.

The top 10 most recurring spam keywords, in order of frequency, are: \textit{win}, \textit{dvd}, \textit{movies}, \textit{giveaway}, \textit{deals}, \textit{horror}, \textit{bluray}, \textit{ebay}, \textit{gameofthrones}, \textit{movie}. Manual inspection of the 87 keywords suggests that three main types of social media spam campaigns occur in this scenario:

\begin{itemize}
\item Tobacco-related product promotions (sales, coupons, discount codes, etc.);
\item Tobacco-unrelated product promotions (sales, coupons, discount codes, etc.), in particular related to entertainment products (dvd, music, books, etc.); 
\item Topic-hijacking campaigns, \textit{i.e.,}  spam  that includes tobacco-related keywords to attract the attention of users to tweets related to completely different topics, including movies and TV shows (keywords like \textit{gameofthrones}, \textit{fiftyshades}, \textit{hungergames}, \textit{celebs}, \textit{ageofultron}, \textit{insurgent}, and many others), and offline news events (e.g., \textit{charlestonshooting}, \textit{ericgarner}). 
\end{itemize}

The phenomenon of Twitter hashtag hijacking has been documented extensively~\cite{chu2012detecting, hadgu2013political, jain2015hashjacker, jackson2015hijacking}. In the following analysis, we do not make a specific distinction between different types of spam campaigns. However, in the future, we will try to determine whether campaign types, as well as different scopes and intents lead to different social spam dynamics.

\begin{figure}[t]
\sidecaption[t]
\includegraphics[width=\columnwidth]{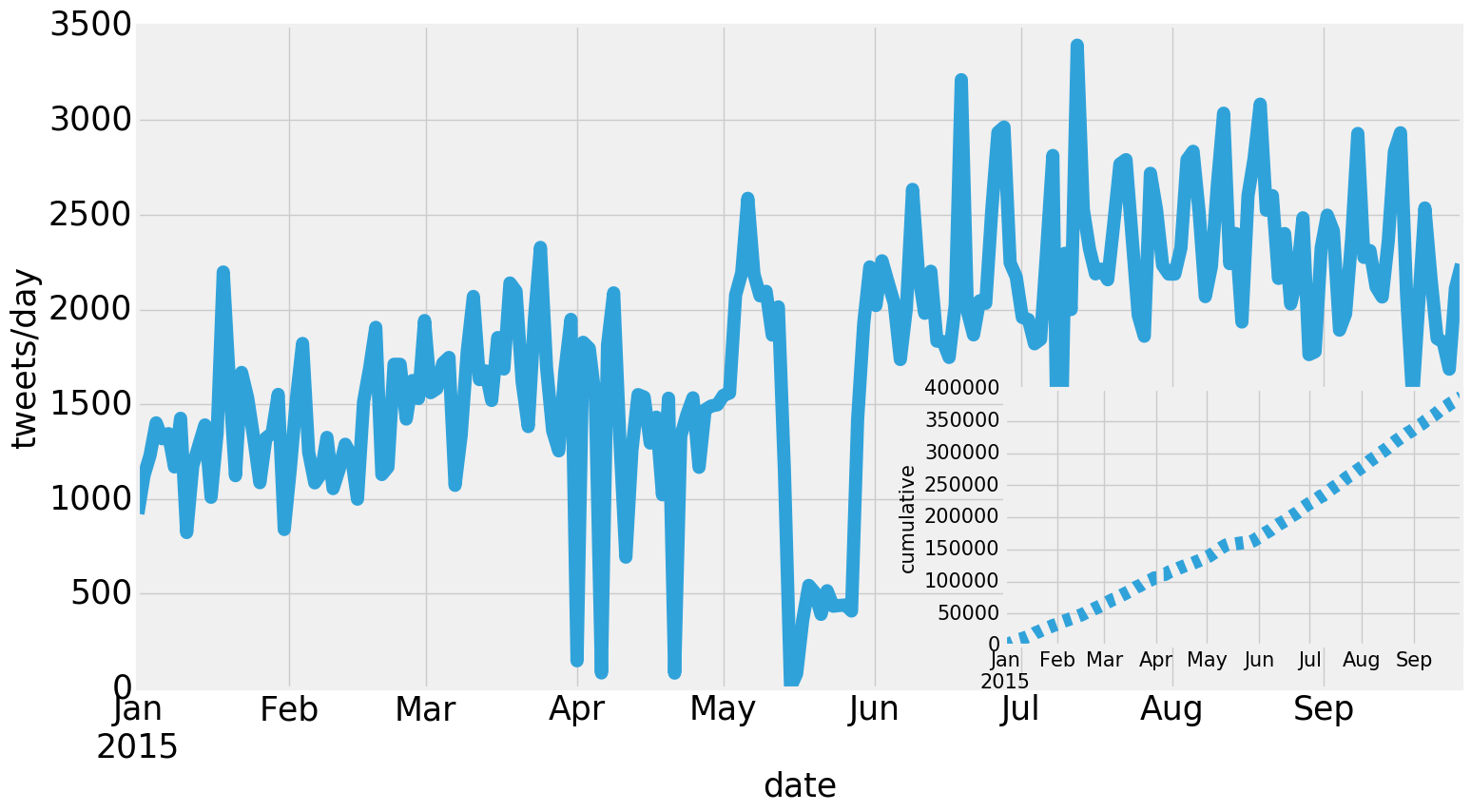}
\caption{Timeline of the volume of spam tweets per day during the observation period. The inset shows the cumulative count. A few drops visible in April and May are associated with Twitter data collection service outages.}
\label{fig:spam-timeline}       
\end{figure}

\subsubsection*{Descriptive Data Statistics}
Our initial exploratory analysis aims at highlighting the temporal dynamics of social spam production. Figure~\ref{fig:spam-timeline} shows the timeline of the volume of spam tweets per day in our dataset. Overall, we can note a mild upward trend over the course of the 9 months of observation. By the end of the year, the volume of tweets per day is roughly twice that of the beginning. This growth suggests the effectiveness of social spam in the tobacco-related context: if ineffective, the cost associated with running social spam campaigns would outweigh their benefits and therefore we would observe declining trends.

\begin{figure}[t]
\sidecaption[t]
\includegraphics[width=\columnwidth]{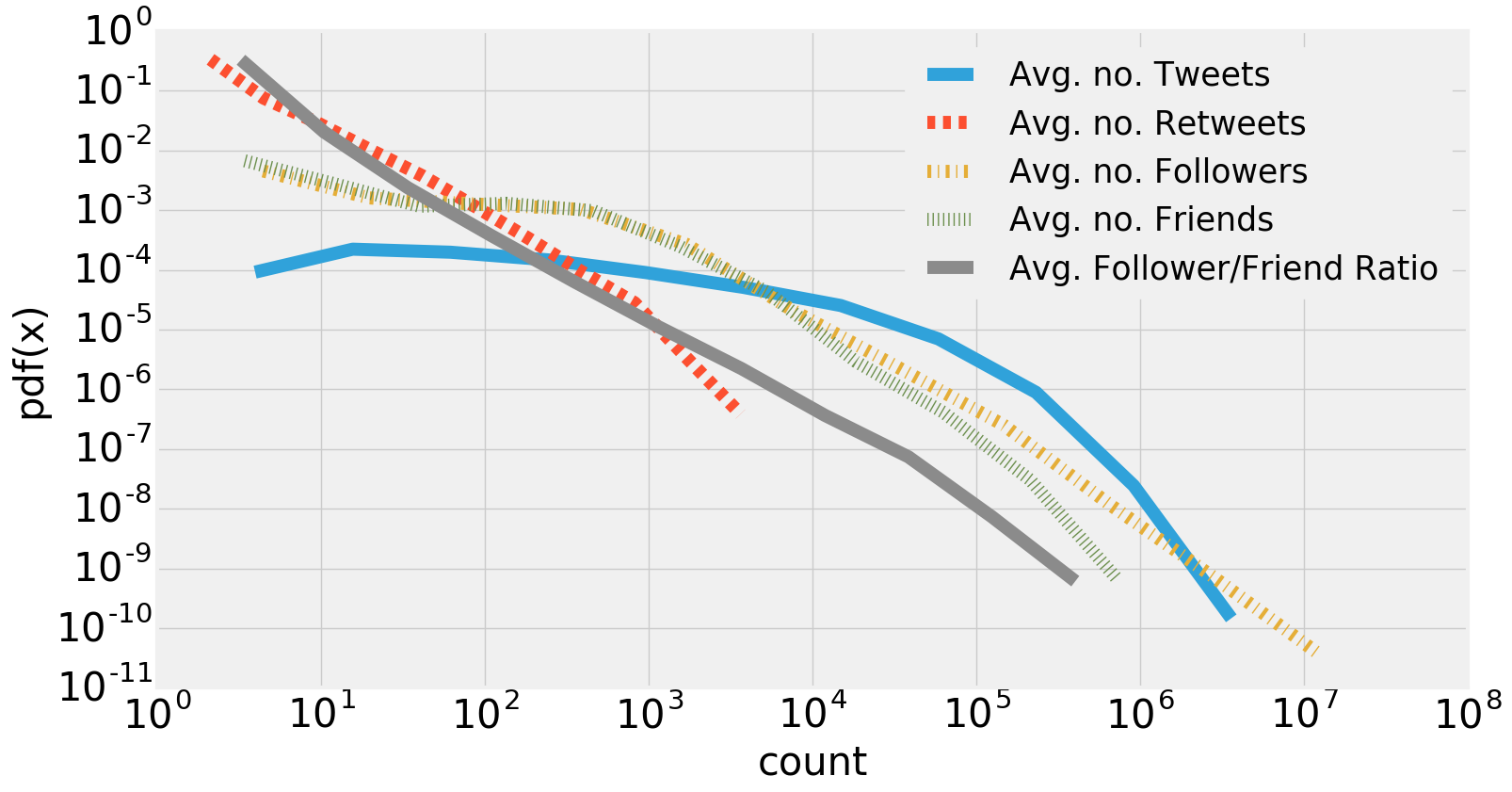}
\caption{Distributions of the average number of tweets, retweets, followers, friends, and follower vs friend ratio of the users in our spam dataset.}
\label{fig:spam-distributions}       
\end{figure}

After assessing that social spam was ``alive and well'' during our analysis period, we moved forward to provide a statistical characterization of the actors therein involved: the Twitter spammers.
Figure~\ref{fig:spam-distributions} shows the distribution of the average number of posted tweets, obtained retweets, number of followers and friends, and follower vs. friend ratio, for the set of users in our spam dataset. The averages are calculated across the 9-month observation period. A few observations are in order. Firstly, although all distributions exhibit the  heavy tails typical of social networks~\cite{barabasi2005origin, ahn2007analysis}, some are significantly different from others. For example, the distribution of posted tweets is somewhat unexpected; if compared with the distribution of obtained retweets, which exhibits the typical power-law like behavior (i.e., a truncated straight line in the log-log plot of Figure~\ref{fig:spam-distributions}), the distribution of posted tweets appear anomalous. In particular, it appears that there is roughly the same probability of observing accounts with a number of posted tweets that spans from a few to over ten thousands: this is represented by the nearly-flat slope of the blue solid curve in the regime $10 \leq x < 10^4$. After that point, the probability decreases very rapidly. This unusual behavior is commonly linked to the activity of social bots. Their activity, however, does not catch up with the lack of influence they are typically characterized by, and therefore the amount of average retweets that most of these accounts receive is orders of magnitude lesser than the amount of tweets they post. Concluding, both the friends and follower distribution exhibit uncommon shapes, suggesting the presence of two different regimes, one for $10 \leq x < 10^3$ and one for $x \geq 10^3$. The slope in the former regime is nearly flat, whereas in the latter both distributions decay with more typical heavy tails suggesting the presence of accounts with a very large number of friends and followers, another interesting behavior associated with two types of users: influential individuals, or social bots.
Next, we study in detail the relation between activity and connectivity patterns.

\begin{figure}[t]
\sidecaption[t]
\includegraphics[width=\columnwidth]{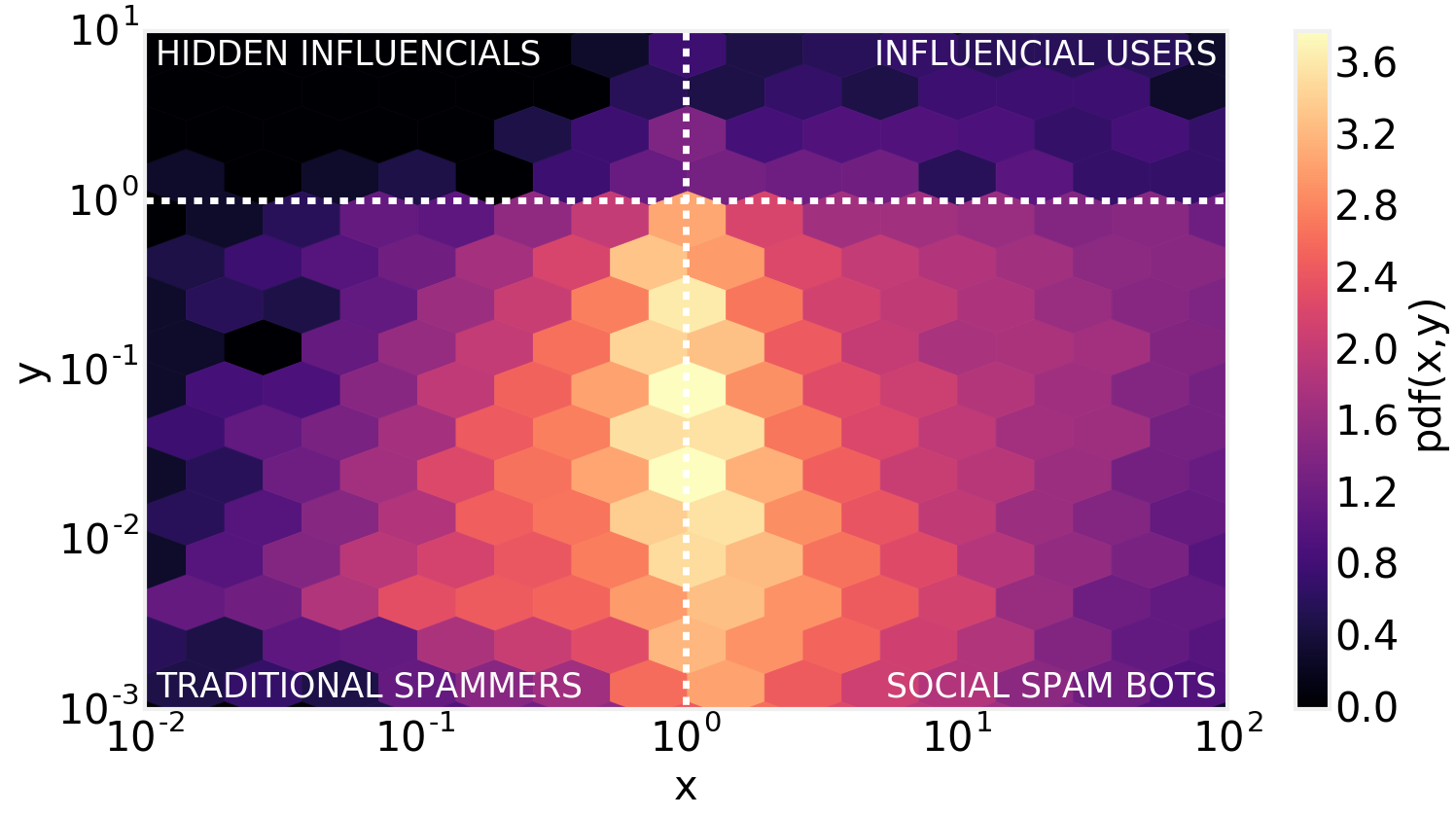}
\caption{Dynamical activity-connectivity map of the users in our dataset. The $x$ axis represents the proportional variation of followers/friends for each user over the accounted time period. The $y$ axis represents the proportional variation of received/posted tweets of each user over the time period.}
\label{fig:spam-DAC}       
\end{figure}

\subsubsection*{Dynamical Activity-Connectivity Maps}
The analysis above was static: taking the average values of the five features above made the results oblivious of the temporal dynamics of activity and connectivity as they unfold over the observation time. 
We now plan to investigate what effect the progression of activity levels of a user has on their connectivity evolution (and viceversa). In Figure~\ref{fig:spam-DAC} we provide a \textit{Dynamical Activity-Connectivity map}: we recently introduced this type of maps~\cite{ferrara2017computational, varol2014evolution} as dynamic variants of the map proposed by Gonzalez-Bailon and collaborators---see Figure 4 in the paper titled \textit{Broadcasters and Hidden Influentials in Online Protest Diffusion}~\cite{gonzalez2013broadcasters}.

Figure~\ref{fig:spam-DAC} shows the probability density of users in the two-dimensional space where the x-axis represents the growth of network connectivity, and the y-axis conveys the messaging activity rate. 
For a given user $u$, $x_u$ and $y_u$ are here defined as 

$$x_u = \frac{1+\delta f_u}{1+\delta F_u} \qquad  \mbox{and} \qquad y_u= \frac{1+\delta rt_u}{1+\delta t_u}.$$

We use the notations $f_u$ and $F_u$ to identify the number of followers and friends, respectively, of a user $u$. 
The variations of followers and friends of user $u$ over a period of time $t$ are thus defined as  $\delta f_u = \frac{f_u^{max} - f_u^{min}}{t}$ and $\delta F_u = \frac{F_u^{max} - F_u^{min}}{t}$; the length of time $t$ is defined as the number of days of $u$'s activity, measured from registration to last observed activity (this varies from user to user).
Finally, the variations of received retweets, and posted tweets, are defined as $\delta rt_u = \frac{rt_u^{max} - rt_u^{min}}{t}$ and $\delta t_u = \frac{t_u^{max} - t_u^{min}}{t}$ respectively, where $rt_u$ and $t_u$ are the number of obtained retweets and posted tweets by user $u$ during the period of activity $t$.
 
All values are added to the unit to avoid zero-divisions and to allow for logarithmic scaling (i.e., in those cases where the variation is zero). The ``heat'' (the color intensity) in the map represents the joint probability density $pdf(x,y)$ for users with given values of $x$ and $y$. The plot also introduce a bin normalization to account for the logarithmic binning. 

The \textit{Dynamical Activity-Connectivity map} we conceived is interpreted as follows: the bulk of the joint probability density mass should be observed in the neighborhood of $(1,1)$, as the majority of accounts would usually exhibit a  comparable variation along the two dimensions. That would be in line with what all previous social media studies where this type of map was employed reported~\cite{gonzalez2013broadcasters, ferrara2017computational, varol2014evolution}. However, the results Figure~\ref{fig:spam-DAC} shows are unprecedented: we hypothesize that this is due to the spam dynamics characterizing this dataset.
Let us discuss the two dimensions of \textit{connectivity growth} and \textit{activity rate} separately.

The \textit{connectivity growth} is captured by the $x$ axis and, in our case, ranges roughly between $10^{-2}$ and $10^2$. 
Users for which $x>1$ (i.e., $10^0$) are those with a followership that grows much faster than the rate at which these users are following others. In other words, they are acquiring social network popularity (followers) at a fast-paced rate. Note that, if a user is acquiring many followers quickly, but s/he is also following many users at a similar rate, the value of $x$ will be near 1. This is a good property of our measure because it is common strategy on social media platforms, especially among bots~\cite{ferrara2016rise,bessi2016social}, to indiscriminately follow others in order to seek for reciprocal followerships. Our Dynamical Activity-Connectivity map will discriminate users with fast-growing followerships, who will appear in the right-hand side of the map, from those who adopt that type of reciprocity-seeking strategy. The former group can be associated with highly popular users with a fast-paced followership growth. 
According to Gonzalez-Bailon and collaborators~\cite{gonzalez2013broadcasters} this category is composed by two groups: \textit{influential users} and \textit{information broadcasters}, depending on their activity rates.
Values of $x<1$ indicate users who follow others at a rate higher than that they are being followed; they fall in the left-hand side of the map.
According to Gonzalez-Bailon and collaborators, these are mostly the \textit{common users}, although the so-called \textit{hidden influentials} also sit in this \textit{low-connectivity} regime.

As for what concerns the $y$ axis, it measures the \textit{activity rate}, i.e.,  the rate at which a user receives retweets versus how frequently s/he tweets.
Users with values of $y>1$ are those who receive systematically more retweets with respect to how frequently they tweet. This group of users  can be referred to as \textit{influentials},  i.e., those who are referred to significantly more frequently than others in the conversation; they fall in the upper region of the map, and according to Gonzalez-Bailon \textit{et al.}, depending on their connectivity growth can be divide in influential ($x>1$) and hidden influential ($x<1$) users.
Conversely, users with values of $y<1$ are those who post exceedingly more tweets than the retweets they receive. This group would generally represents the common-user behavior ($x<1$), although information broadcasters ($x>1$) also exhibit the same \textit{low-activity} rate. These users fall in the lower region of the map.

Now that a reading of dynamical activity-connectivity maps has been provided, we can proceed with interpreting Figure~\ref{fig:spam-DAC}: the bottom-left quadrant reports the most common users, those with both activity and connectivity growth lesser than $1$. In our case, we identify these accounts as traditional spammers. Manual validation of some of these accounts revealed that they employ simple automatic posting strategies, thus they generate a very large number of tweets, but they never attract other users' attention and thus they are rarely retweeted. We identified over 27K such accounts.

Conversely, the upper-right quadrant reports users with the higher connectivity growth and activity rates. These are influential accounts:  they systematically attract other users' attention by receiving lots of retweets compared with how often they tweet, and their followerships grow at a very fast pace. Influential users are quite rare in this context, and in fact we identified only 438 users according to our method. Manual inspection of all these users revealed that our technique correctly detects influential users which are not bots: accounts in this category include official accounts of movies and TV shows (e.g., \textit{Avengers}, \textit{CaptainAmerica}, \textit{Divergent}, \textit{GameOfThrones}, etc.), and various official accounts of tobacco-related sellers.

\begin{figure}[t]
\sidecaption[t]
\includegraphics[width=\columnwidth]{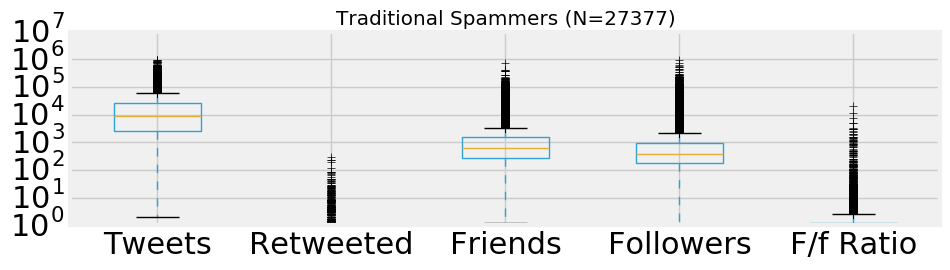}
\includegraphics[width=\columnwidth]{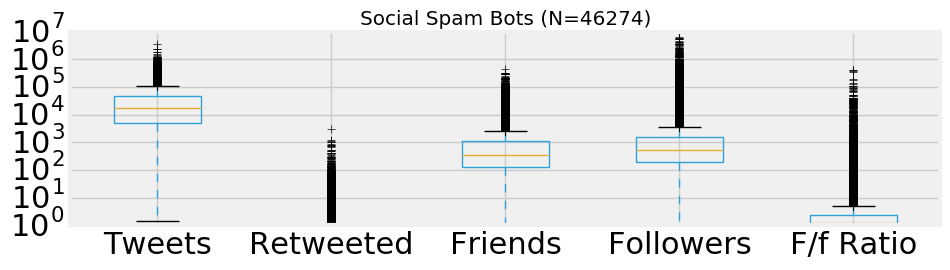}
\includegraphics[width=\columnwidth]{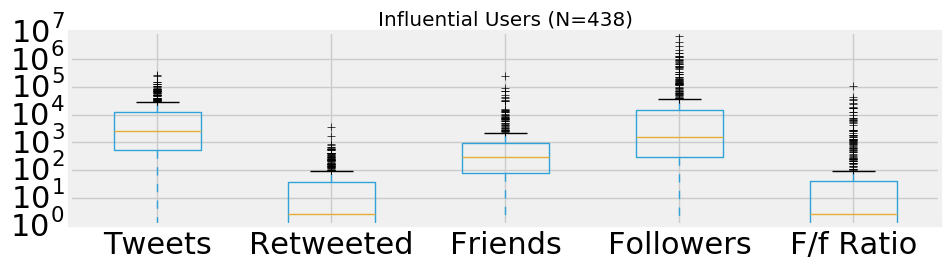}
\caption{Box plot of the distributions of posted tweets, obtained retweets, number of friends and followers, and follower/friend ratio for the main three classes of users in our spam dataset.}
\label{fig:221}       
\end{figure}

Lastly, social spam bots sit in the bottom-right quadrant. Differently from traditional spammers, their connectivity growth is much more similar to that of influential accounts. Their followership increases at a pace higher than their following others. They still produce disproportionately more tweets than the retweets they receive, but their embeddedness in the social network looks somewhat effective. Further analysis reveals that many of these spam bots tend to reciprocate followership to external users (accounts not present in the spam dataset) but also tend to follow each other; this coordinated behavior gives the appearance of network influence. We identified over 46K social spammers, the majority class by far in our spam dataset. Finally, we detected only 47 hidden influentials, too few to warrant further analysis.

Figure~\ref{fig:221} provides a different view on the five features characterizing the users in the three classes. As opposed to spammers, influential users receive significantly more attention (retweets), significantly more followers than friends (thus a much higher followers/friends ratio), and on average post one order of magnitude fewer tweets than bots. Concluding, the only significant difference between traditional spammers and social spam bots is their social network: social bots exhibit more followers than friends on average; the viceversa is true for traditional spam bots.

\begin{figure}[t]
\sidecaption[t]
\includegraphics[width=\columnwidth]{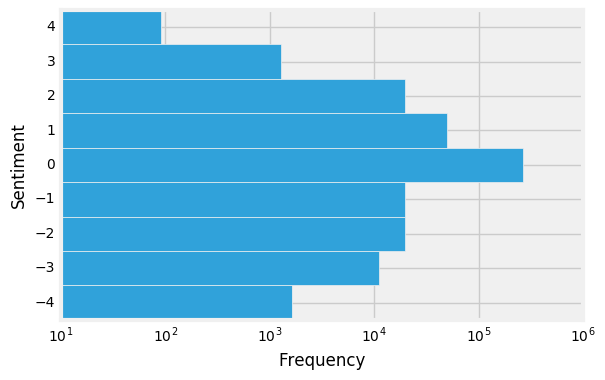}
\caption{Distribution of tweet sentiment scores (SentiStrength) in the spam dataset.}
\label{fig:232}       
\end{figure}

\subsubsection*{The Interplay between Sentiment of Spam Bots}
We conclude our analysis with a high-level investigation of the interplay between spam sentiment and spam bot characteristics. We applied the same Sentiment Analysis technique, i.e., \textit{SentiStrength}, as in the previous case study, to our spam dataset. Figure~\ref{fig:232} shows the distribution of sentiment scores for the tweets in our corpus. The distribution exhibits its typical peak around zero~\cite{thelwall2013heart, ferrara2015quantifying}. However, in contrast with respect to previous findings on Twitter sentiment obtained using SentiStrength~\cite{ferrara2015quantifying}, the distribution in the spam dataset appears skewed toward negativeness. In particular, roughly one order of magnitude more strongly negative tweets ($S\leq -3$) appear than strongly positive ones ($S\geq 3$).

Worth noting, this dataset is significantly smaller and topically biased (i.e., it covers only spam) than the comprehensive Twitter dataset we previously studied~\cite{ferrara2015quantifying}: we hypothesize that some correlation may exist between this atypical sentiment distribution and the role of spam bots. 

\begin{figure}[t]
\sidecaption[t]
\includegraphics[width=.49\columnwidth]{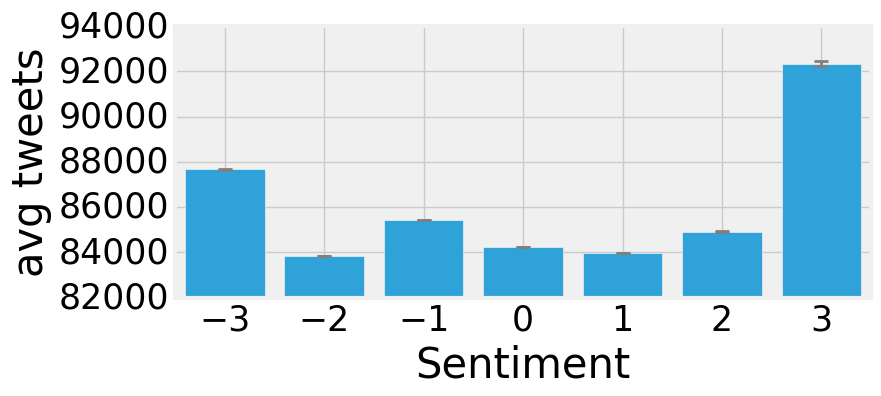}
\includegraphics[width=.49\columnwidth]{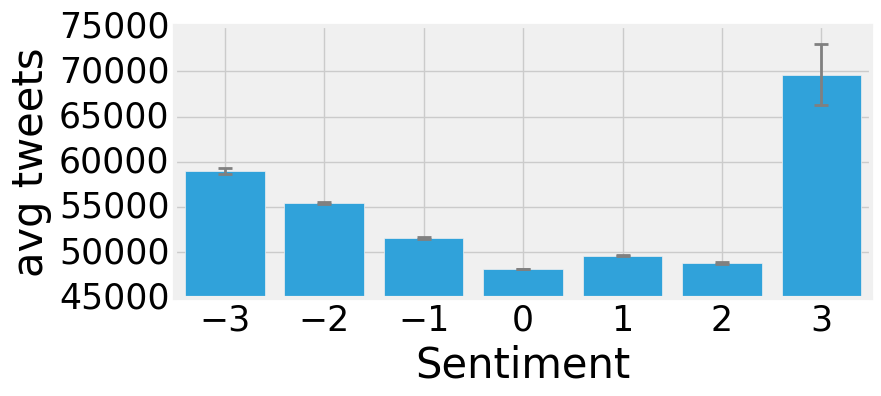}
\caption{Average number of  tweets posted as a function of tweet's sentiment, calculated only on tweets retweeted at most once (left) and on those that have been retweeted more than once (right).}
\label{fig:2a}       
\end{figure}

\begin{figure}[t]
\sidecaption[t]
\includegraphics[width=.49\columnwidth]{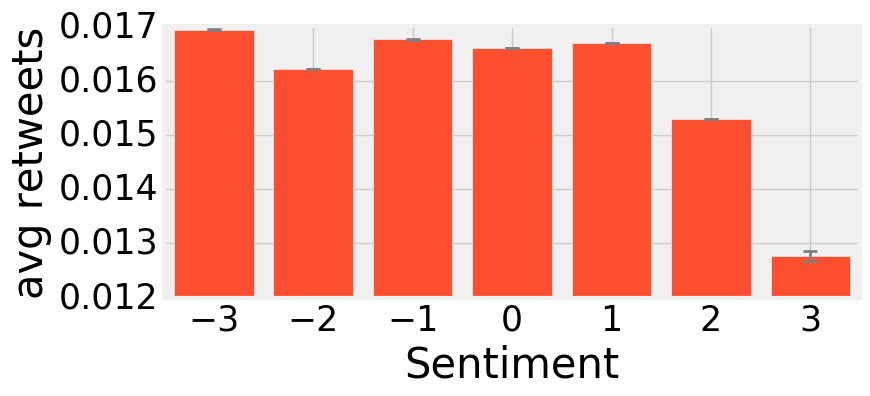}
\includegraphics[width=.49\columnwidth]{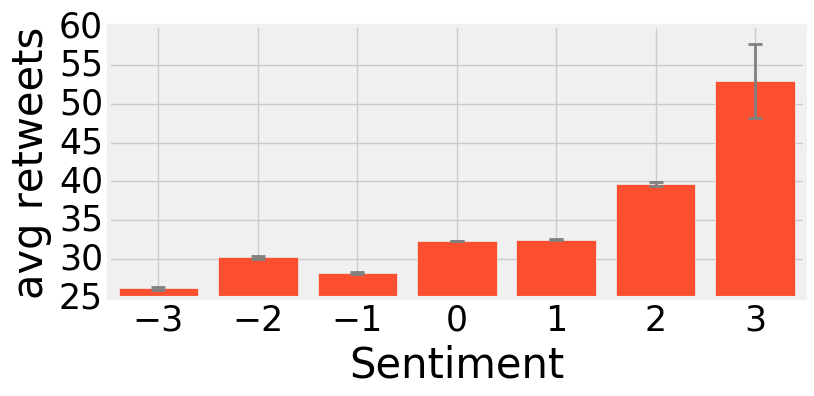}
\caption{Average number of obtained retweets as a function of sentiment,calculated only on tweets retweeted at most once (left) and on those that have been retweeted more than once (right).}
\label{fig:2b}       
\end{figure}

\begin{figure}[t]
\sidecaption[t]
\includegraphics[width=.49\columnwidth]{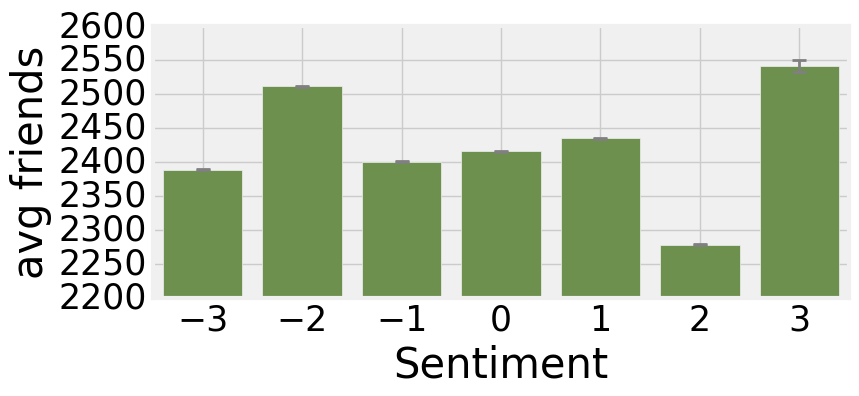}
\includegraphics[width=.49\columnwidth]{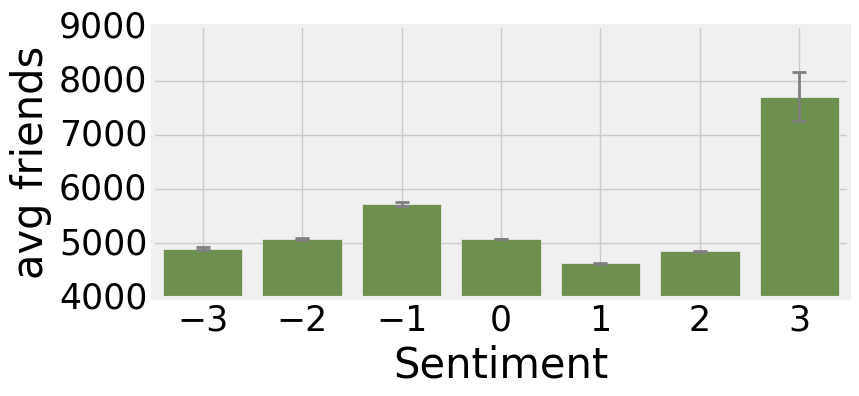}
\caption{Average number of user friends as a function of sentiment, calculated only on tweets retweeted at most once (left) and on those that have been retweeted more than once (right).}
\label{fig:2d}       
\end{figure}

To this purpose, in Figures~\ref{fig:2a}--\ref{fig:2d} we plotted four features we used to characterize the bots (i.e., \textit{number of posted tweets, obtained retweets, friends, and followers}). All figures report error bars (obtain hardly noticeable) that convey the standard error of the sampled average feature distributions.
We will use them for diagnostic purpose, i.e., to highlight anomalies in spam dynamics with respect to organic social media sentiment~\cite{ferrara2015quantifying}.
Given the exiguous number of tweets with extremely positive or negative sentiment (i.e., $S=4$ or $S=-4$), next we will limit our analysis to values of sentiment in the range $-3 \leq S \leq 3$.

\begin{figure}[t]
\sidecaption[t]
\includegraphics[width=.49\columnwidth]{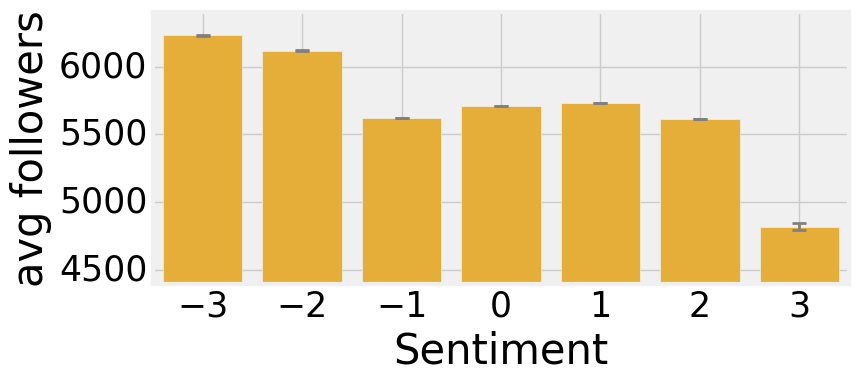}
\includegraphics[width=.49\columnwidth]{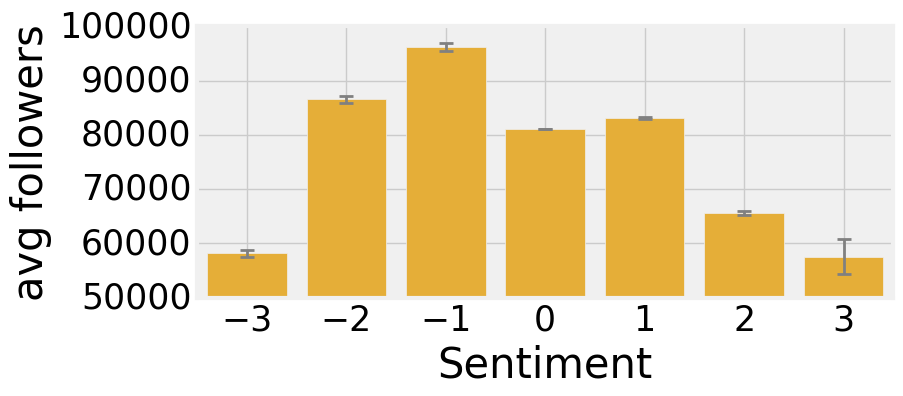}
\caption{Average number of user followers as a function of sentiment, calculated only on tweets retweeted at most once (left) and on those that have been retweeted more than once (right).}
\label{fig:2c}       
\end{figure}



The interpretations of the bar plots in Figures~\ref{fig:2a}--\ref{fig:2d} is the following: given a fixed value of sentiment $x$, then $y$ is the average value of the selected feature for all tweets with sentiment equal to $x$. 
Plots on the left are for the subset of tweets retweeted at most once; plots on the right are for tweets retweeted more than once. The separation is carried out to address the issue of activity heterogeneity highlighted before (\textit{cf.} Figure~\ref{fig:spam-distributions}) and is necessary to avoid problems like the \textit{Simpson Paradox}~\cite{wagner1982simpson}.

For sake of example, let us discuss the left panel of Figure~\ref{fig:2a} that  shows the distribution of the \textit{average number of tweets} posted by users, which were retweeted at most once, as a function of sentiment. 

Let us consider sentiment $S=3$ (there are about 1,300 such tweets in our dataset, \textit{cf.} Figure~\ref{fig:232}): the average number of tweets posted by the users who posted one such tweet with sentiment $S=3$ is about 92K. This is significantly higher than for every other sentiment score, denoting the fact that users who post strongly positive tweets (e.g., promotional tweets) on average posted significantly more tweets than the others. It is also worth noting that an average value of tweets nearing the hundred of thousands clearly denotes very highly-active accounts, and likely some form of automatic posting---a common feature of spam bots. 

The right panel of Figure~\ref{fig:2a} shows how this pattern is preserved even for the set of tweets that have been retweeted more than once: moreover, the distribution takes a U-like shape, suggesting that also accounts that post negative tweets exhibit much more activity than average. This suggests that some spam campaigns may not be necessarily positive. Indeed, if one compares this result with the previous case study on the manipulation of political campaigns, some interesting similarities emerge. In other words, spam at times can aim to smear some products, e.g., those from competitors.

Figure~\ref{fig:2b} shows another interesting patterns. The left panel again captures tweets that have been retweeted at most once; the right panel captures more popular tweets and exhibits a striking difference if compared to the left one: increasingly positive sentiment yields significantly more retweets. This is known as \textit{positivity bias}, i.e., the emergence of a strong preference for retweeting positive messages; such bias was already observed in our prior Twitter analysis~\cite{ferrara2015quantifying}. Strongly positive tweets obtain on average more than twice the number of retweets than negative or neutral ones. It is worth hypothesizing that, in the spam scenario, this pattern may also conceal some form of coordinated activity, i.e., bots may retweet other bots' spam in an orchestrated fashion. 

Further clues supporting this hypothesis come from Figure~\ref{fig:2d}, in particular the right panel:  users associated with positive tweets that are retweeted very often, all exhibit a number of friends that is nearly twice as much as others. Inspecting users who follow on average over 7K accounts  revealed strong reciprocity---another very common bot characteristic highlighted multiple times above. 

Looking at the complementary picture, i.e. the distribution of followers reported in Figure~\ref{fig:2c}, reinforces our hypothesis: left and right panels illustrate two very different scenarios, with the latter showing how users who post very positive or very negative tweets attracted significantly fewer followers than others: bots involved in spam campaigns do not commonly exhibit large followership (\textit{cf.} Figure~\ref{fig:spam-DAC}).

Concluding, our diagnostics revealed characteristic patterns that may conceal clues to decode the strategies employed by spam bots to spread the content they produce, and try giving spam a legitimate appearance.

\section{Conclusions}
Social bots have become a pervasive presence in social media platforms. 
Applications of social bots have been documented in a variety of scenarios, including for public opinion manipulation and for social spam campaigns. The focus of this chapter was to investigate both these domains, and in particular to study the interplay between bots and information diffusion in the two scenarios.

In Section~\S\ref{sec:bots}, we reviewed how social bots are created, and how they operate in social media platforms. We also briefly discussed the challenges of, and the methods to detecting them, covering techniques based on graph-centric detection, crowdsourcing, and traditional feature-based supervised learning. 

Section~\S\ref{sec:election} presented our first case study, discussing how social bots have been used during the 2016 US Presidential Election to sway the conversation around the presidential candidates. In this section  we revised in detail the tools we used for social bot detection, namely \textit{Bot Or Not}, for Sentiment Analysis, namely \textit{SentiStrength}, and for partisanship detection. 

We also summarized the results of our study on political manipulation~\cite{bessi2016social}, providing in particular two data-driven insights: first, we noted that social bots generate as much engagement, at least in terms of obtained retweets, than humans, suggesting the fact that humans cannot tell apart bots from other humans very easily when rebroadcasting politics-related information on Twitter. Second, we illustrated the interplay between content sentiment and social bots, highlighting a few partisanship differences (e.g., Trump bots single-handedly generated the most positive supporting content of their candidate in the entire analyzed dataset).

Finally, in Section~\S\ref{sec:spam} we proposed a second case study, and new results and analyses about the effects of social spam bots on the diffusion of social spam campaigns within the tobacco-related conversation on Twitter. First, we identified the presence of three types of spam campaigns: \textit{(i)} relative to tobacco products; \textit{(ii)} relative to products unrelated to the tobacco industry, e.g., entertainment products; and, finally, \textit{(iii)} instances of topic hijacking, namely the use of hashtags and keywords related to the tobacco industry to attract individuals' attention on issues completely unrelated to that, e.g., social issues connected to news events in the offline world. 

By means of a newly-introduced method named \textit{Dynamical Activity-Connectivity map}, we also revealed the existence of different classes of spam accounts, including traditional spammers and social spam bots; we also discussed a statistical characterization of their most typical features. In conclusion, we provided an analysis of the interplay between sentiment and spam bots, revealing patterns that may conceal strategies of bot coordination, and the resulting effects in terms of spam diffusion.

Our findings in both case studies exemplify the potential for social media abuse: whether at stakes is the right to exercise unbiased elections and therefore  democracy itself, or the exposure to illegitimate spam and propaganda, social media manipulation can have devastating societal effects. This study encourages future  efforts of the research community to address the various facets of this form of abuse.

%
%
\bibliographystyle{abbrv}
\bibliography{ferrara.bib}

\end{document}